\documentstyle[epsf,preprint,eqsecnum,aps]{revtex}

\begin{document}
\draft
\preprint{KYUSHU-HET 28}
\title{Mesons in the massive Schwinger model\\on the light-cone}
\author{Koji Harada$^{1,2}$
\footnote{electronic address: koji1scp@mbox.nc.kyushu-u.ac.jp},
Atsushi Okazaki$^1$
\footnote{electronic address: zaki1scp@mbox.nc.kyushu-u.ac.jp}, and
Masa-aki Taniguchi$^1$
\footnote{electronic address: mass1scp@mbox.nc.kyushu-u.ac.jp}}
\address{Department of Physics, Kyushu University\\Fukuoka 812-81,
JAPAN$^1$}
\address{Department of Physics, The Ohio-State University\\
Columbus, Ohio 43210, USA$^2$}
\date{September 25, 1995}
\maketitle

\begin{abstract}
We investigate mesons in the bosonized
massive Schwinger model in
the light-front Tamm-Dancoff approximation in the strong coupling
region.
We confirm
that the three-meson bound state has a few percent fermion six-body
component
in the strong coupling region when expressed in terms of fermion
variables,
consistent with our previous calculations.
We also discuss some qualitative features of the three-meson bound
state
based on the information about the wave function.
\end{abstract}
\narrowtext
%
%
\section{Introduction}
Recently there has been increasing interest in
light-front field theory\cite{reviews}. In particular the light-front
Tamm-Dancoff (LFTD) approximation\cite{PHW} has proven to be very
powerful,
as a nonperturbative approach to the relativistic bound state
problem.
It has been successfully applied to two-dimensional
models\cite{twodim}
as well as to four-dimensional Yukawa theory\cite{Yukawa}.
It is important to note that this new
approach not only reproduced known results correctly, but also
brought us
new results, which have never been obtained by other
methods\cite{HSTY,SMY}.
(See also Refs.~\cite{dlcq} for some of such ``new'' results in the
discretized light-cone quantization (DLCQ) approach\cite{pb}.)

In a previous paper\cite{HOTsix}, we studied the massive Schwinger
model in
the framework of the LFTD approximation up to
including fermion six-body states. We showed that
 (1) the two-meson bound state has a negligibly small six-body
component, (2) the three-meson bound state does exist,
and (3) the two-meson bound state is well described in
terms of the wave function of the relative motion.
Surprisingly, however, the
six-body component of the three-meson bound state is quite small,
though it
is large in comparison with those of
other states below the three-meson threshold.
Typically
it is at most a few percent for a small value of the fermion mass.
Despite this small
six-body component, we identified it as the three-meson bound state
based on the
following reasons:
(1) Since the meson creation operator contains
the fermion annihilation operators (see Ref.~\cite{HOTsix} for the
notation),
\begin{eqnarray}
A^{\dag}(p)&=&\int^{p}_0{dk\over(2\pi)\sqrt{k(p-k)}}\psi(k,p-k)
b^{\dag}(k)d^{\dag}(p-k)\nonumber\\
&+&\int^{\infty}_0{dk\over(2\pi)\sqrt{k(p+k)}}\varphi(p+k,k)
[b^{\dag}(p+k)b(k)-d^{\dag}(p+k)d(k)],\label{Amassive}
\end{eqnarray}
the three-meson bound state ($\sim
{A^\dagger}^3\left\vert0\right\rangle$)
naturally contains fermion two- and four-body components,
beside the six-body component.
(2) This state is charge conjugation odd, while a two-meson state
should be
charge conjugation even.

Unfortunately, we were not able to {\em explain} why
the six-body component of the
three-meson bound state is so small. It is not the aim of this paper
to
do it, but to further confirm our conclusion and to give some
arguments
by using bosonization.

How can we further justify the identification despite the smallness
of the six-body component of the state?
As emphasized in the previous paper, we have a simple picture of the
massive
Schwinger model in the strong coupling region. Because the massless
theory
(the strong coupling limit) is a free massive boson (meson) theory,
we expect
that it becomes a weakly interacting massive boson theory
once a small mass term is
included (the strong coupling region). In addition, light-front field
theory
provides us with the simple vacuum. Actually, these two
allowed us to describe the
two-meson bound state in terms of the wave function of the relative
motion\cite{HOTsix}.
One may think, therefore, of constructing the wave function
of the relative motion
for the three-meson bound state in a similar way. If such a
description
is a good approximation of the state, one may justify that it is a
three-meson state. It turns out, however,
that it is almost infeasible in terms of the fermion variables
because
it is difficult to find a simple set of basis functions which
satisfies
all the symmetry properties.

A simpler way we will follow in this paper is to consider
the bosonized theory.

Is the Tamm-Dancoff approximation good for the bosonized theory too?
The meson has an internal structure in the sense
that the wave function, $\psi(k,p-k)$ in eq.~(\ref{Amassive}),
has nontrivial momentum dependence, which becomes
negligible in the strong coupling limit, i.e.,
the meson becomes ``structureless.''
The Fock states in the bosonized
theory is that of this ``structureless'' meson.
The ``structureless'' meson state is a good
approximation in the strong coupling region. It means
that the Tamm-Dancoff approximation is
good for the bosonized theory in the strong coupling region.
(The wave function gets the momentum dependence from many-body
states.)
In the bosonized theory, it is easy to find a set of basis functions
which
satisfies all the symmetry properties.

In this paper, we consider the bosonized massive Schwinger model in
the
LFTD approximation up to including three-boson states. We show that
almost all of our previous calculations\cite{HOTsix} are consistent
with the
results obtained in the bosonized theory in the strong coupling
region.
In particular, we show that
the three-meson bound state, which is almost 100\% a
three-boson state, has a few percent {\it fermion\/} six-body
component.

We are aware of the limitation of our analysis in this paper. The
limitation
comes from (1) the normal-ordering problem in light-front field
theory, and
(2) the unboundedness of the Hamiltonian for the charge conjugation
even
sector. We think that we can avoid these problems if we are confined
in the
very strong coupling region. We will discuss these problems in the
final
section.
%
%
%
\section{Bosonized massive Schwinger model}
\subsection{LFTD for the bosonized massive Schwinger model}
It is well known that the massive Schwinger model\cite{CJS,C}
($QED_{1+1}$
with massive fermion),
\begin{eqnarray}
{\cal L}&=&-{1\over4}F_{\mu\nu}F^{\mu\nu}
+\bar{\psi}[\gamma^\mu(i\partial_{\mu}-eA_{\mu})
-m]\psi\label{lagrangian}\ ,\\
&&F_{\mu\nu}=\partial_{\mu}A_{\nu}-\partial_{\nu}A_{\mu}\ ,\nonumber
\end{eqnarray}
has an equivalent bosonic form
\begin{equation}
{\cal L}_b={1\over2}\partial_{\mu}\phi\partial^{\mu}\phi
-{\mu^2\over2}\phi^2
+cm\mu: \cos(2\sqrt{\pi}\phi):\ ,\label{bosonizedlagrangian}
\end{equation}
where $c$ is numerical constant ($c=e^{\gamma}/2\pi$, with $\gamma$
being
the Euler constant), and $\mu=e/\sqrt{\pi}$.
In the following we set $\mu=1$, so the strong coupling region
corresponds
to small fermion masses. We quantize this
model on the light cone.

In the equal-time quantization, the normal-ordering is well-defined
in the
interaction picture. (The above normal-ordering is with respect to
the free
boson with mass $\mu$.) It is known, on the other hand, that a naive
normal-ordering fails in the light-front quantization, because one
usually
neglects generalized tadpole diagrams\cite{griffin}. The effects of
the
generalized tadpoles amount to the renormalization of the coupling
constant\cite{burkardt}.
In this paper, however, we do not consider this coupling constant
renormalization, simply assuming that the effects of it are
negligibly
small for the very strong coupling region. We will discuss it in the
final section.

Expanding $\phi$ (in the Schr\"odinger picture)
in terms of
creation and annihilation operators,
\begin{eqnarray}
&&\phi(x^-)=\int_0^{\infty}{dp^+\over2\sqrt{\pi}p^+}
[a(p^+)e^{-ip^+x^-}
+a^{\dag}(p^+)e^{ip^+x^-}]\ ,\\
&&[a(p^+),a^{\dag}(q^+)]=p^+\delta(p^+-q^+)\ ,\ [a(p^+),a(q^+)]=
[a^{\dag}(p^+),a^{\dag}(q^+)]=0\ ,
\end{eqnarray}
we obtain the light-cone Hamiltonian:
\begin{eqnarray}
&&P^-=H=H_0+V\ ,\nonumber\\
&&H_0={\tilde{\mu}^2\over2}\int_0^{\infty}{dp\over
p^2}a^{\dag}(p)a(p)\ ,\\
&&V=V_4+V_6+\cdots\ ,\nonumber
\end{eqnarray}
where $\tilde{\mu}^2=1+4\pi cm$. Note that because
we are going to consider
the Tamm-Dancoff (TD) truncation up to including three-boson states,
the
interaction terms containing more than six creation and/or
annihilation
operators are irrelevant, and will be ignored hereafter. The
interaction
terms are expressed in terms of $a^\dagger$ and $a$ in the following
way,
\begin{eqnarray}
V_4&=&-{cm\over4!}(2\pi)\int_0^{\infty}\prod_{i=1}^4{dp_i\over p_i}
[4\delta(p_1+p_2+p_3-p_4)a^{\dag}_1a^{\dag}_2a^{\dag}_3a_4\nonumber\\
&&+6\delta(p_1+p_2-p_3-p_4)a^{\dag}_1a^{\dag}_2a_3a_4
+4\delta(p_1-p_2-p_3-p_4)a^{\dag}_1a_2a_3a_4]\ ,\\
V_6&=&{cm\over6!}(2\pi)\int_0^{\infty}\prod_{i=1}^6{dp_i\over p_i}
[6\delta(p_1+p_2+p_3+p_4+p_5-p_6)
a^{\dag}_1a^{\dag}_2a^{\dag}_3a^{\dag}_4
a^{\dag}_5a_6\nonumber\\
&&+15\delta(p_1+p_2+p_3+p_4-p_5-p_6)
a^{\dag}_1a^{\dag}_2a^{\dag}_3a^{\dag}_4
a_5a_6\nonumber\\
&&+20\delta(p_1+p_2+p_3-p_4-p_5-p_6)
a^{\dag}_1a^{\dag}_2a^{\dag}_3a_4a_5a_6
\nonumber\\
&&+15\delta(p_1+p_2-p_3-p_4-p_5-p_6)
a^{\dag}_1a^{\dag}_2a_3a_4a_5a_6\nonumber\\
&&+6\delta(p_1-p_2-p_3-p_4-p_5-p_6)
a^{\dag}_1a_2a_3a_4a_5a_6]\ ,
\end{eqnarray}
where we use the abbreviation $a_i=a(p_i)$.

The spectrum of the model is obtained by solving
the Einstein-Schr\"odinger equation,
$2P^+P^-\left\vert\phi\right\rangle=M^2\left\vert\phi\right\rangle$.
We truncate the Fock space up to including three-boson states.
\begin{eqnarray}
&&\left\vert\phi\right\rangle_{\cal
P}=\left\vert\phi_1\right\rangle_{\cal P}+
\left\vert\phi_2\right\rangle_{\cal
P}+\left\vert\phi_3\right\rangle_{\cal P}
\ ,\\
&&\left\vert\phi_1\right\rangle_{\cal P}=\phi_1a^{\dag}({\cal
P})\left\vert0
\right\rangle\ ,\nonumber\\
&&\left\vert\phi_2\right\rangle_{\cal P}={1\over
\sqrt{2}}\int_0^{\cal P}
\prod_{i=1}^2{dp_i\over\sqrt{p_i}}\delta(\sum_{i=1}^2p_i-{\cal P})
\phi_2(p_1,p_2)a^{\dag}(p_1)a^{\dag}(p_2)\left\vert0\right\rangle\ ,
\nonumber\\
&&\left\vert\phi_3\right\rangle_{\cal P}={1\over
\sqrt{3!}}\int_0^{\cal P}
\prod_{i=1}^3{dp_i\over\sqrt{p_i}}\delta(\sum_{i=1}^3p_i-{\cal P})
\phi_3(p_1,p_2,p_3)
a^{\dag}(p_1)a^{\dag}(p_2)a^{\dag}(p_3)\left\vert0\right
\rangle\ ,\nonumber
\end{eqnarray}
where $\cal P$ is the eigenvalue of the momentum operator $P^+$.
Note that $\phi_2(p_1,p_2)$
and $\phi_3(p_1,p_2,p_3)$ are totally symmetric functions of the
arguments.

The boson creation operator can be expressed in terms of the fermion
and antifermion creation and annihilation operators (see
Ref.~\cite{HOTsix}
for the notation),
\begin{eqnarray}
a^{\dag}(p)&=&\int^{p}_0{dk\over(2\pi)\sqrt{k(p-k)}}
b^{\dag}(k)d^{\dag}(p-k)\nonumber\\
&+&\int^{\infty}_0{dk\over(2\pi)\sqrt{k(p+k)}}
[b^{\dag}(p+k)b(k)-d^{\dag}(p+k)d(k)].\label{Amassless}
\end{eqnarray}
It is clear that the symmetry under $\phi \to -\phi$ (or $a^\dagger
\to -a^\dagger$) is the charge conjugation symmetry of the fermionic
theory.
Because of this symmetry, a state has a definite transformation
property,
\begin{equation}
\left\vert\phi\right\rangle_{\cal P}^{e}\to\left\vert\phi\right
\rangle_{\cal P}^{e}\ ,
\quad \left\vert\phi\right\rangle_{\cal
P}^{o}\to-\left\vert\phi\right
\rangle_{\cal P}^{o}\ .
\end{equation}
It is easy to see that $\left\vert\phi_1\right\rangle_{\cal P}$
and $\left\vert\phi_3\right\rangle_{\cal P}$ are odd while
$\left\vert\phi_2\right\rangle_{\cal P}$ is even.

We rescale momenta as $p_i\to x_i=p_i/{\cal P}$, and the wave
functions,
$\phi_2(p_1,p_2)$ and $\phi_3(p_1,p_2,p_3)$, are replaced by
$\phi_2(x_1,x_2)$
and $\phi_3(x_1,x_2,x_3)/\sqrt{{\cal P}}$, respectively.
The Einstein-Schr\"odinger
equation leads to two sets of eigenvalue equations for the wave
functions,
according to the transformation property under the charge
conjugation.
The even one involves only $\phi_2$:
\begin{equation}
M^2\phi_2(x_1,x_2)=\tilde{\mu}^2\left({1\over x_1}+{1\over
x_2}\right)
\phi_2(x_1,x_2)-cm(2\pi)\int_0^1{dy_1dy_2\over\sqrt{y_1y_2}}
\delta(y_1+y_2-1)
{\phi_2(y_1,y_2)\over\sqrt{x_1x_2}}\ ,
\label{even}
\end{equation}
with $x_1+x_2=1$, while the odd one is the coupled equations
for $\phi_1$ and $\phi_3$:
\begin{eqnarray}
M^2\phi_1&=&\tilde{\mu}^2\phi_1-cm(2\pi){2\over
\sqrt{3!}}\int_0^1\prod_{i=1}^3
{dy_i\over \sqrt{y_i}}\delta(\sum_{i=1}^3y_i-1)\phi_3(y_1,y_2,y_3)\ ,
\label{odd1}\\
M^2\phi_3(x_1,x_2,x_3)&=&
\tilde{\mu}^2\sum_{i=1}^3{1\over x_i}\phi_3(x_1,x_2,x_3)\nonumber\\
&&-cm(2\pi)\int_0^1{dy_1dy_2\over\sqrt{y_1y_2}}[\delta(y_1+y_2+x_1-1)
{\phi_3(y_1,y_2,x_1)\over\sqrt{x_2x_3}}\nonumber\\
&&{}+\delta(y_1+y_2+x_2-1){\phi_3(y_1,y_2,x_2)\over\sqrt{x_3x_1}}
+\delta(y_1+y_2+x_3-1){\phi_3(y_1,y_2,x_3)\over\sqrt{x_1x_2}}]
\nonumber\\
&&{}+cm(2\pi){2\over3!}\int_0^1\prod_{i=1}^3{dy_i\over\sqrt{y_i}}
\delta(\sum_{i=1}^3y_i-1){\phi_3(y_1,y_2,y_3)\over\sqrt{x_1x_2x_3}}
\nonumber\\
&&{}-cm(2\pi){2\over\sqrt{3!}}{\phi_1\over \sqrt{x_1x_2x_3}}\ ,
\label{odd2}
\end{eqnarray}
with $x_1+x_2+x_3=1$.
Note that because the even sector (\ref{even}) does not depend on
$V_6$, the Hamiltonian is not bounded from below. We will see,
however, that we get reasonable results if we do not employ
many basis functions
(see below) and are confined in the very strong coupling region.

These complicated equations are converted to two matrix eigenvalue
equations
by expanding the wave functions in terms of basis functions. The
choice of
the basis functions is very important for efficient numerical work.
We choose the following basis functions so that
we can calculate the matrix elements analytically.
\begin{eqnarray}
\phi_2(x_1,x_2)&=&\sum_{l}b_l(x_1x_2)^{l+1/2}\ ,\label{twobasis}\\
\phi_3(x_1,x_2,x_3)&=&\sum_{\bf l}c_{\bf l}(x_1x_2x_3)^{l_1+1/2}
(x_1^{l_2}+x_2^{l_2}+x_3^{l_2})\ \label{threebasis},
\end{eqnarray}
where $l=0, 1, 2, \cdots, N_1$, $l_1=0, 1, 2, \cdots, N_2$, and
$l_2=0, 2, \cdots, 2N_3$.
It is easy to see that any symmetric polynomial
in $x_1,\ x_2,\ x_3$
with the constraint $x_1+x_2+x_3=1$ can be expressed by using the
above basis functions
(up to $(x_1x_2x_3)^{1/2}$)
(\ref{threebasis}).
%
%
\subsection{States in terms of fermion variables}
How are those states expressed in terms of fermion operators?
By using
(\ref{Amassless}), it is straightforward to express
$\left\vert\phi_2\right\rangle_{\cal P}$ as follows,
\begin{eqnarray}
&&\left\vert\phi_2\right\rangle_{\cal P}=\left\vert\psi_2^{(2)}\right
\rangle_{\cal P}+\left\vert\psi_4^{(2)}\right\rangle_{\cal P}\
,\nonumber\\
&&\left\vert\psi_2^{(2)}\right\rangle_{\cal P}=\int_0^{\cal P}
{dk_1dk_2\over2\pi\sqrt{k_1k_2}}\delta(k_1+k_2-{\cal
P})\psi_2^{(2)}(k_1,k_2)
b^{\dag}_1d^{\dag}_2\left\vert0\right\rangle ,\\
&&\left\vert\psi_4^{(2)}\right\rangle_{\cal P}={1\over2}\int_0^{\cal
P}
\prod_{i=1}^4{dk_i\over\sqrt{2\pi k_i}}\delta(\sum_{i=1}^4k_i-{\cal
P})
\psi_4^{(2)}(k_1,k_2;k_3,k_4)
b^{\dag}_1b^{\dag}_2d^{\dag}_3d^{\dag}_4\left
\vert0\right\rangle\ ,\nonumber
\end{eqnarray}
where
\begin{eqnarray}
\psi_2^{(2)}(k_1,k_2)&=&{1\over\sqrt{2}}\left[\int_0^{k_1}dq
{\phi_2(k_1-q,k_2+q)\over\sqrt{(k_1-q)(k_2+q)}}-(k_1\leftrightarrow
k_2)\right]
\ ,\\
\psi_4^{(2)}(k_1,k_2;k_3,k_4)&=&-{1\over\sqrt{2}}
\left[{\phi_2(k_1+k_3,k_2+k_4)
\over\sqrt{(k_1+k_3)(k_2+k_4)}}-(k_1\leftrightarrow k_2)\right]\ .
\end{eqnarray}
In a similar way, $\left\vert\phi_3\right\rangle_{\cal P}$ is
expressed as
\begin{eqnarray}
&&\left\vert\phi_3\right\rangle_{\cal P}=\left\vert\psi_2^{(3)}\right
\rangle_{\cal P}+\left\vert\psi_4^{(3)}\right\rangle_{\cal P}
+\left\vert\psi_6^{(3)}\right\rangle_{\cal P}\ ,\nonumber\\
&&\left\vert\psi_2^{(3)}\right\rangle_{\cal P}=\int_0^{\cal P}
{dk_1dk_2\over2\pi\sqrt{k_1k_2}}\delta(k_1+k_2-{\cal
P})\psi_2^{(3)}(k_1,k_2)
b^{\dag}_1d^{\dag}_2\left\vert0\right\rangle ,\\
&&\left\vert\psi_4^{(3)}\right\rangle_{\cal P}={1\over2}\int_0^{\cal
P}
\prod_{i=1}^4{dk_i\over\sqrt{2\pi k_i}}\delta(\sum_{i=1}^4k_i-{\cal
P})
\psi_4^{(3)}(k_1,k_2;k_3,k_4)
b^{\dag}_1b^{\dag}_2d^{\dag}_3d^{\dag}_4\left
\vert0\right\rangle\ ,\nonumber\\
&&\left\vert\psi_6^{(3)}\right\rangle_{\cal P}={1\over3!}\int_0^{\cal
P}
\prod_{i=1}^6{dk_i\over\sqrt{2\pi k_i}}\delta(\sum_{i=1}^6k_i-{\cal
P})
\psi_6^{(3)}(k_1,k_2,k_3;k_4,k_5,k_6)
b^{\dag}_1b^{\dag}_2b^{\dag}_3d^{\dag}_4
d^{\dag}_5d^{\dag}_6\left\vert0\right\rangle\ ,\nonumber
\end{eqnarray}
where
\begin{eqnarray}
\psi_2^{(3)}(k_1,k_2)&=&{1\over\sqrt{3!}}\left[\left(\int_0^{k_1}dl_1
\int_0^{l_1}dl_2{\phi_3(k_1-l_1,l_1-l_2,k_2+l_2)\over
\sqrt{(k_1-l_1)(l_1-l_2)(k_2+l_2)}}\right.\right.
\nonumber\\
&&{}-\left.\left.\int_0^{k_1}dl_1
\int_0^{k_2}dl_2{\phi_3(k_1-l_1,k_2-l_2,l_1+l_2)
\over\sqrt{(k_1-l_1)(k_2-l_2)(l_1+l_2)}}\right)+(k_1\leftrightarrow
k_2)
\right]\ ,\\
\psi_4^{(3)}(k_1,k_2;k_3,k_4)&=&-{\sqrt{3!}\over4}
\Big[\int_{k_3}^{k_1}dl
{\phi_3(l,k_1+k_3-l,k_2+k_4)\over\sqrt{l(k_1+k_3-l)(k_2+k_4)}}
-(k_1\leftrightarrow k_2)-(k_3\leftrightarrow k_4) \\
&&\quad{}+(k_1\leftrightarrow k_2,k_3\leftrightarrow k_4)\Big]\
,\nonumber
\end{eqnarray}
and
\begin{eqnarray}
\lefteqn{\psi_6^{(3)}(k_1,k_2,k_3;k_4,k_5,k_6)}\nonumber \\
&&=-{1\over\sqrt{3!}}\left[\left(
{\phi_3(k_1+k_4,k_2+k_5,k_3+k_6)
\over\sqrt{(k_1+k_4)(k_2+k_5)(k_3+k_6)}}
-(k_2\leftrightarrow k_3)\right)+\left(\begin{array}{c}k_1,k_2,k_3\\
\mbox{cyclic}\end{array}\right)\right]\ .
\end{eqnarray}

The even state $\left\vert\phi\right\rangle^e_{\cal P}$
contains the fermion two- and four-body components:
\begin{eqnarray}
{}^e_{\cal P'\ }\!\!\!\left\langle\phi\vert\phi\right\rangle^e_{\cal
P}
&=&{\cal P}\delta({\cal P'}-{\cal P})\left[W_2^e+W_4^e\right]\
,\nonumber\\
W_2^e&=&\int_0^1dx_1dx_2\delta(x_1+x_2-1)
\vert\psi_2^{(2)}(x_1,x_2)\vert^2\ ,
\label{componentsym}\\
W_4^e&=&\int_0^1\prod_{i=1}^4dx_i\delta(\sum_{i=1}^4x_i-1)\vert
\psi_4^{(2)}(x_1,x_2;x_3,x_4)\vert^2\ ,\nonumber
\end{eqnarray}
and the odd state $\left\vert\phi\right\rangle^o_{\cal P}$
contains the fermion two-, four- and six-body components:
\begin{eqnarray}
{}^o_{\cal P'\ }\!\!\!\left\langle\phi\vert\phi\right\rangle^o_{\cal
P}
&=&{\cal P}\delta({\cal P'}-{\cal P})\left[W_2^o+W_4^o+W_6^o\right]\
,
\nonumber\\
W_2^o&=&\vert\phi_1\vert^2+\int_0^1dx_1dx_2\delta(x_1+x_2-1)\vert
\psi_2^{(3)}(x_1,x_2)\vert^2\ ,\label{componentant}\\
W_4^o&=&\int_0^1\prod_{i=1}^4dx_i\delta(\sum_{i=1}^4x_i-1)\vert
\psi_4^{(3)}(x_1,x_2;x_3,x_4)\vert^2\ ,\nonumber\\
W_6^o&=&\int_0^1\prod_{i=1}^6dx_i\delta(\sum_{i=1}^6x_i-1)\vert
\psi_6^{(3)}(x_1,x_2,x_3;x_4,x_5,x_6)\vert^2\ .\nonumber
\end{eqnarray}

Once we numerically obtain the eigenstates of the
Einstein-Schr\"odinger equation, i.e.,
the basis function expansion coefficients of the boson wave
functions,
(\ref{twobasis}) and (\ref{threebasis}), we
can calculate all of these integrals analytically so that we can
obtain
fermion wave functions. This virtue comes from our clever choice of
the
simple set of basis functions.

We normalize states in a  Lorentz invariant way,
${}_{\cal P'\ }\!\!\!\left\langle\phi\vert\phi\right\rangle_{\cal P}
={\cal P}\delta({\cal P'}-{\cal P})$, so that $W^e_2+W^e_4=
W^o_2+W^o_4+W^o_6=1$. By saying that a state has a $50.000\%$
fermion two-body component, we mean that the state has $W_2=0.50000$.
%
%
\subsection{Numerical Results}\label{num}
\subsubsection{mass spectrum}
\begin{figure}[h]
\vspace{10mm}
\epsfysize=0.8\vsize
\epsfbox{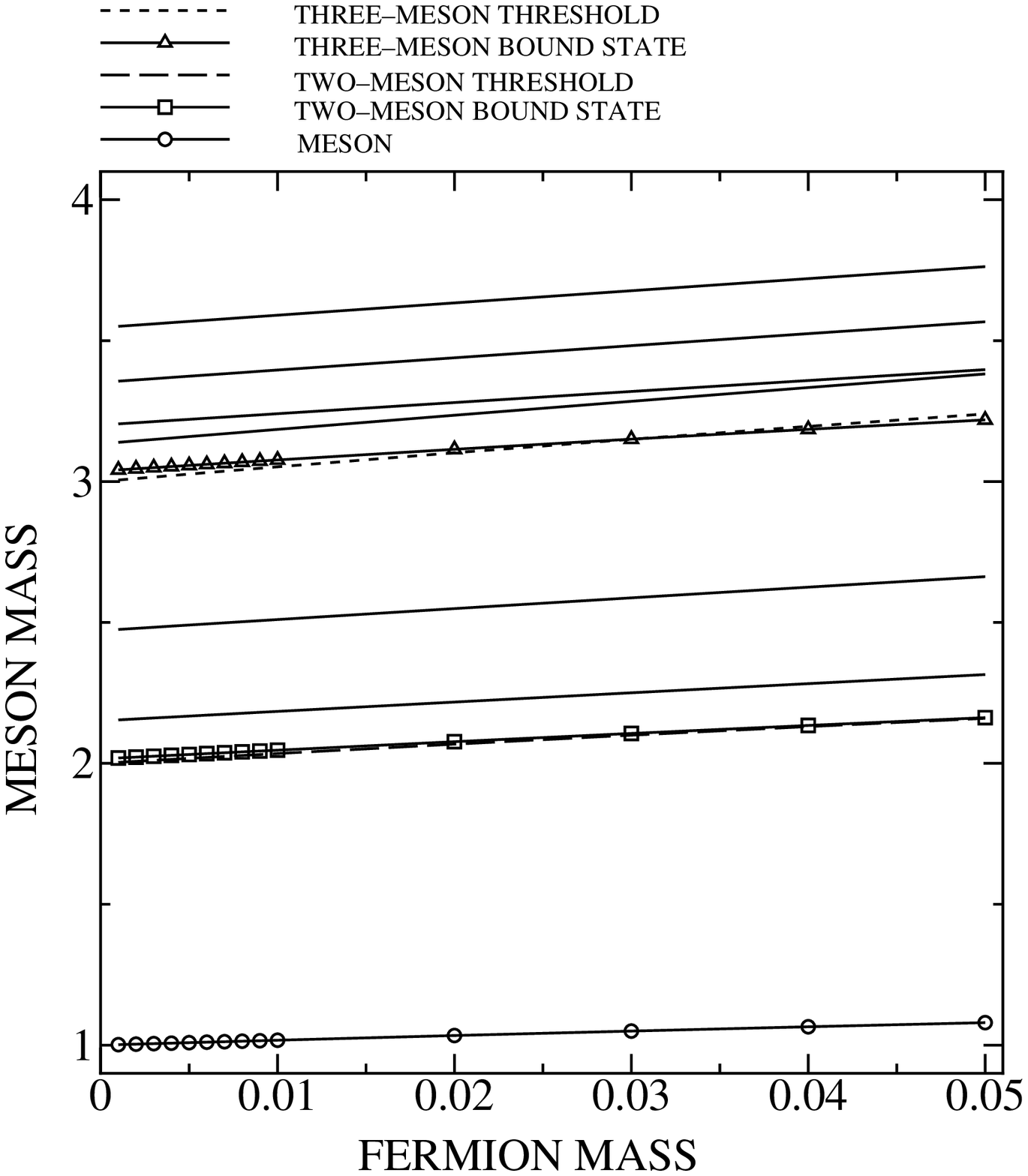}
\vspace{15mm}
\caption{The mass spectrum for $0.001\leq m\leq 0.05$, at $N_1=5,
N_2=4,$ and
$N_3=1$. The dashed and dotted line stand for the two-meson
and three-meson thresholds respectively.}
\label{masses}
\end{figure}

We calculate the invariant masses $M$ for various values of the
fermion mass $m$. We find that convergence is good enough for $N_1=5,
N_2=4,$
and $N_3=1$ (total number of basis functions is $16$).
The mass spectrum for $0.001\leq m\leq 0.05$ is
shown in Fig.~\ref{masses}.
In the following, we concentrate on the case $m=0.01$.

The lowest state is the meson state. It is charge conjugation
odd. It has $99.997\%$ one-boson component and
$0.003\%$ three-boson component. Its mass is $1.0174$.

The second lowest one should be the two-meson bound state, though its
mass appears slightly above the threshold, $M=2.0468$. It is charge
conjugation even. Because of the present approximation,
its two-boson component is $100\%$.
It may contain a four-boson component if we include four-boson
states, but
we expect that it is negligibly small in the strong coupling region.

The third and fourth states come from the charge conjugation even
sector.
These are regarded as two-meson scattering states. They are
completely
two-boson states because of the present approximation.

The fifth one is the three-meson bound state, though its mass also
appears
slightly above the threshold, $M=3.0766$. It is charge conjugation
odd.
It has $0.000\%$ one-boson component and $100.000\%$ three-boson
component.

The two-meson and three-meson bound states appear
below the thresholds when the fermion mass gets larger.
In the strong coupling region, they should appear just below
the threshold. (In the strong coupling limit,
they are just on the thresholds, not being bound.) It is
therefore difficult to get them below the thresholds numerically.

\subsubsection{Fermion components}
How do these states look like if they are expressed in terms of the
fermion variables? By using
(\ref{componentsym}) and (\ref{componentant}), we get the following
results:
(1) The meson state is perfectly a fermion two-body state
($100.00\%$).
(2) The two-meson bound state contains
$45.92\%$ fermion two-body component and $54.08\%$ fermion
four-body component. (Its fermion six-body component is $0\%$ due to
the
present approximation.)
(3) The three-meson bound state, which is of our main
interest, contains $28.38\%$ fermion two-body component and $63.87\%$
fermion four-body component and $7.75\%$ fermion six-body component.

{}For the comparison, we give our previous results\cite{HOTsix}:
(1) The meson
state has $M=1.01813$ and is perfectly a fermion two-body state
($100.000\%$).
(2) The two-meson bound state has $M=2.05612$. It has $54.408\%$
fermion
two-body component and $45.592\%$ fermion four-body component,
with little fermion six-body component.
(3) The three-meson
bound state has $M=3.10814$. It has $44.285\%$ fermion two-body
component,
$53.123\%$ fermion four-body component,  and $2.592\%$ fermion
six-body component.
Obviously these are consistent with the results obtained by using
bosonization, except for the differences in the fermion two-body and
four-body components of the three-meson bound state.

The most important result is the demonstration that the three-meson
bound
state, which is almost perfectly ($100.000\%$) a three-boson state,
has only $7.75\%$ fermion six-body component.
We think that this is a very strong support that the identification
we made
in the previous paper is correct.

\subsubsection{Wave functions}
\begin{figure}
\epsfysize=0.37\vsize
\epsfbox{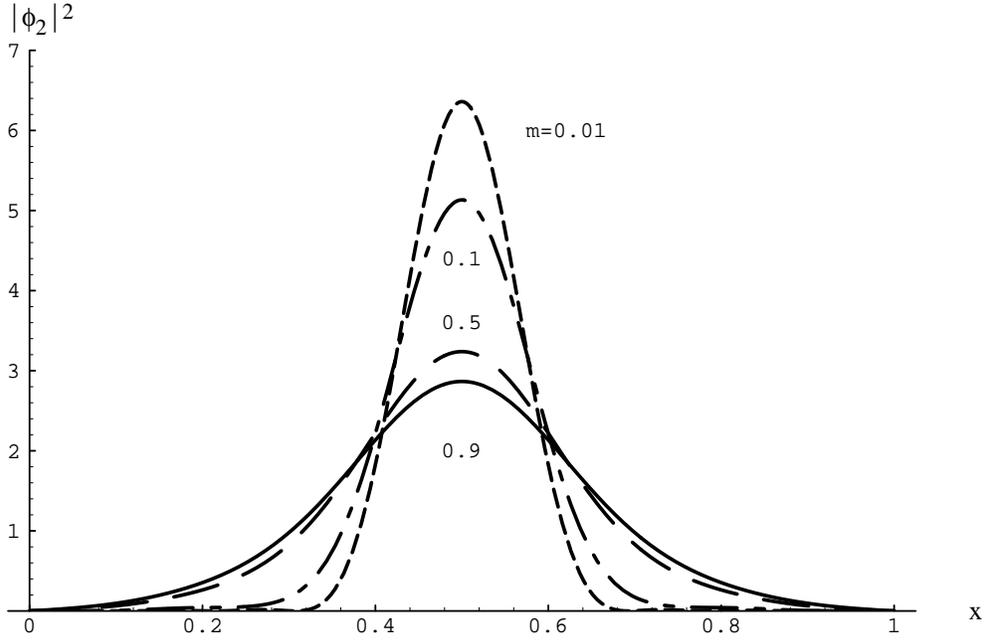}
\vspace{5mm}
\caption{Squared wave function of the relative motion of
the two-meson bound state, $|\phi_2(x,1-x)|^2$, is shown for various
values of the fermion mass. This should be compared with Fig.~6 of
our previous work.}
\label{relative}
\end{figure}

\begin{figure}
\epsfysize=0.37\vsize
\epsfbox{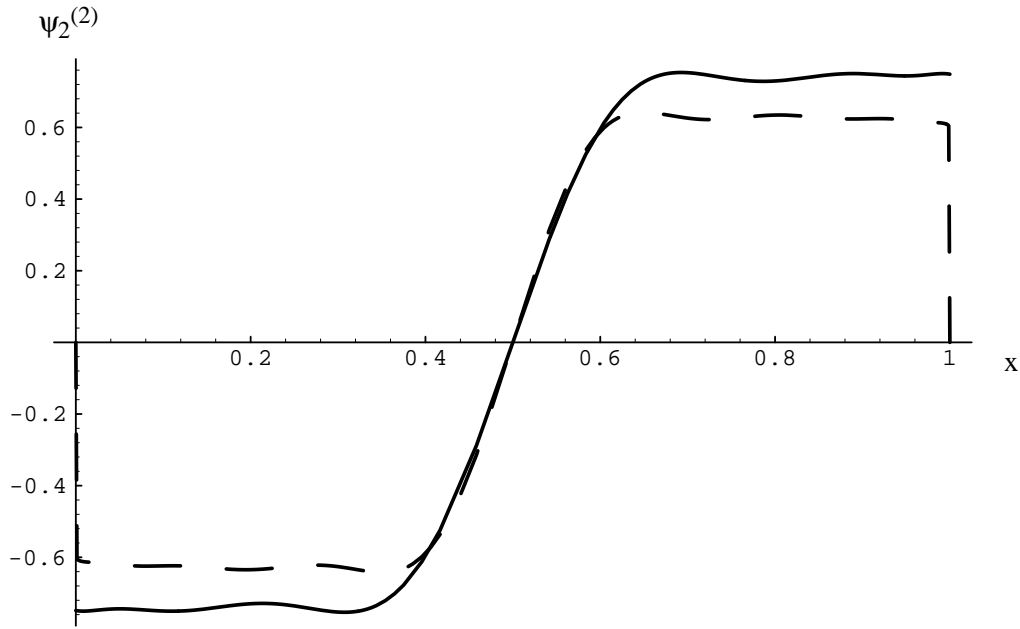}
\caption{Fermion two-body wave function for the two-meson bound
state, $\psi_2^{(2)}(x,1-x)$, is shown (solid line) with that
obtained by the previous calculation in terms of fermion variables
(dashed line) for $m=0.01$. }\label{twowf}
\end{figure}

It is interesting to see the wave function of the ``relative motion''
of
the two-meson bound state (Fig.~\ref{relative}). It looks very
similar to
the one we obtained in the previous paper (Fig.~6 of
Ref.~\cite{HOTsix}).
We can also calculate the fermion two-body wave function of the
two-meson
bound state $\psi^{(2)}_2$ (Fig.~\ref{twowf}) which should be
compared
with Fig.~7 of Ref.~\cite{HOTsix}.
The results are completely consistent with the previous calculations.

\begin{figure}
\epsfysize=0.42\vsize
\epsfbox{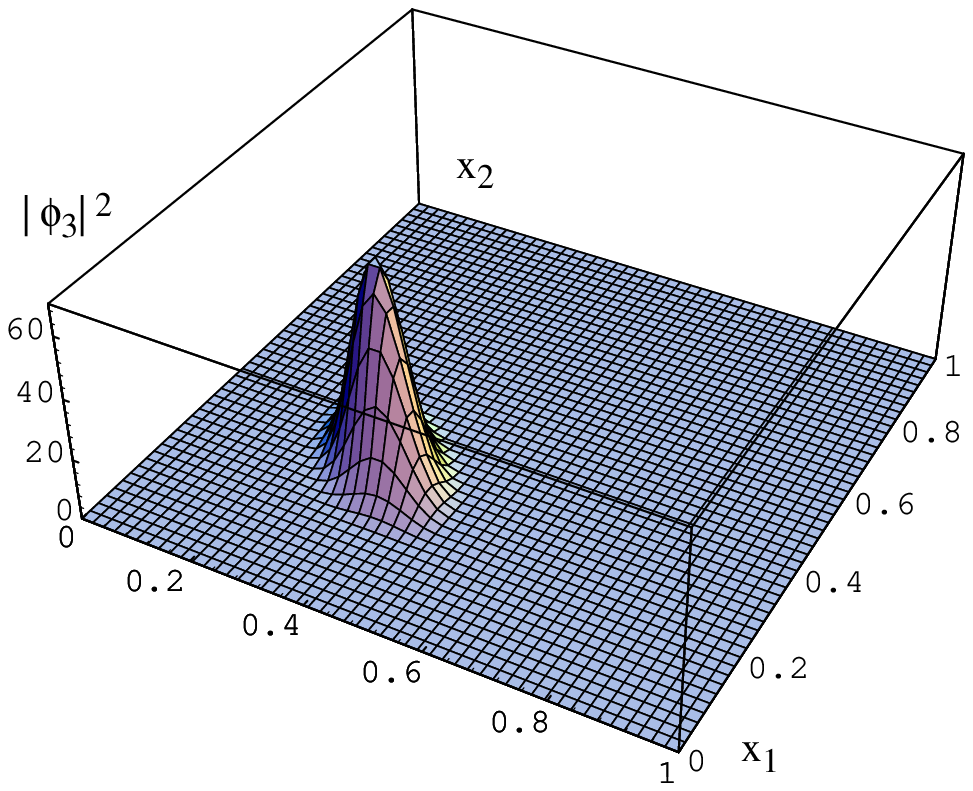}
\caption{Squared wave function of the three-meson bound state,
$\vert\phi_3(x_1,x_2,x_3)\vert^2$, for $m=0.001$ projected on the
$x_1$-$x_2$ plane. Only the region ($x_1>0,\ x_2>0,\ x_1+x_2<1$)
is the support.}
\label{wavefunction0.001}
\end{figure}

\begin{figure}
\epsfysize=0.42\vsize
\epsfbox{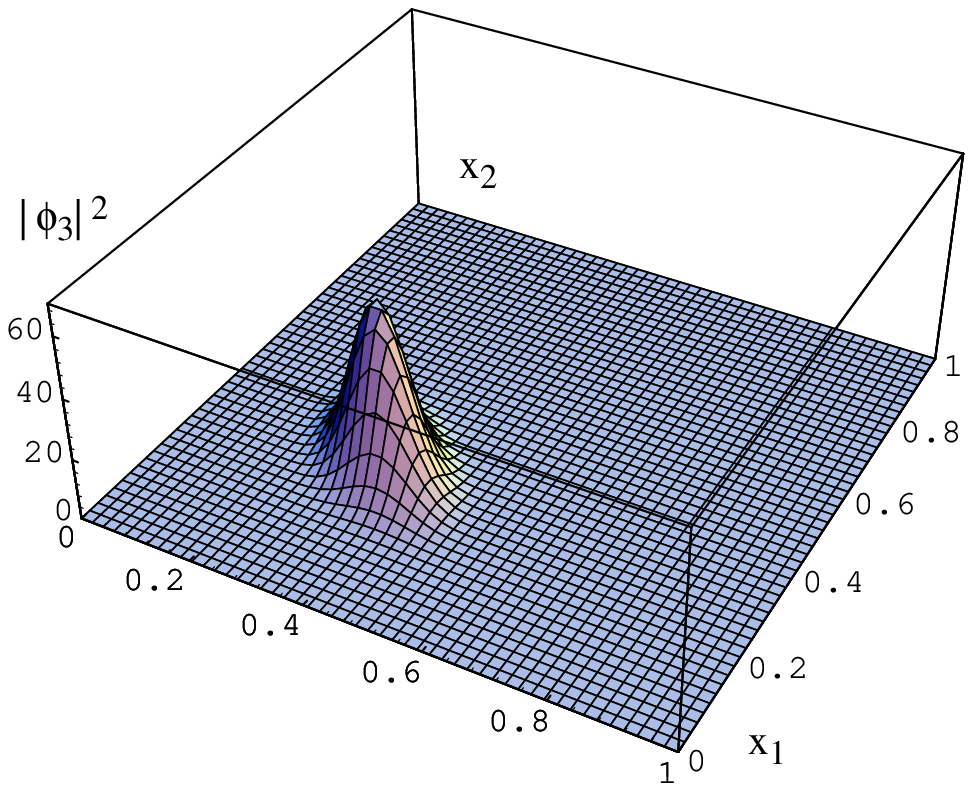}
\caption{Squared wave function of the three-meson bound state,
$\vert\phi_3(x_1,x_2,x_3)\vert^2$, for $m=0.05$
projected on the $x_1$-$x_2$ plane.
Only the region ($x_1>0,\ x_2>0,\ x_1+x_2<1$)
is the support.}
\label{wavefunction0.05}
\end{figure}

We next try to figure out how the three-meson bound state looks
like by examining the wave function. We show $|\phi_3|^2$ of the
three-meson bound state in  Fig.~\ref{wavefunction0.001} (for
$m=0.001$) and Fig.~\ref{wavefunction0.05}
(for $m=0.05$).
Because the wave function in momentum space is spread
for larger fermion masses,
and has a sharp peak for smaller fermion masses, we have an intuitive
picture that for strong couplings the three-meson bound state is
loosely
bound while for weak couplings it has a relatively compact form.
We look for asymmetry which indicates that two of the three mesons
are more closely bound than the third, but we are not able to find
any.
%
%
\section{Summary and Discussions}\label{discussions}
We calculate the mass spectrum of the bosonized massive Schwinger
model
by using the LFTD approximation up to including three-boson states.
We showed that the three-meson bound state, which is almost $100\%$
a three-boson state, has only a few percent fermion six-body
component,
and the result is consistent with our previous LFTD calculations in
terms of
the fermion variables. We also show that the other quantities are
also
consistent with the previous calculations. By obtaining the wave
function
of the three-meson bound state we are able to have an intuitive
picture
of the three-meson bound state, namely, it is loosely bound for
strong
couplings while it has a relatively compact shape  for small
couplings.

In our present approximation, the two-meson bound state cannot have
a non-zero fermion six-body component. Of course, if we include
four-boson
states, it can have a non-zero six-body component. We however expect
that
it will be negligibly small.

The LFTD approximation for the bosonized theory is good only for
strong couplings. There are three reasons: (1)
The Fock space of the bosonized theory is that of a structureless
boson,
so that we neglect the internal structure of the meson as the
first approximation. As is
known from previous calculations, it is not a good description of the
meson to ignore the internal structure already at $m=0.1$.
Note that the nontrivial momentum dependence
of the fermion two-body wave function of the meson, $\psi_2$, comes
from
the many-boson components, when expanded in terms of fermion
variables.
(2) The Hamiltonian for the charge conjugation even sector is not
bounded from
below. (Of course this is because of the Tamm-Dancoff truncation.)
Because of that we observed instabilities when we increase the number
of
basis functions or when we increase the fermion mass. For small
number of
basis functions and small values of the fermion mass, our method does
not
scan the `high energy' states and the low energy states are
insensitive to
the unboundedness of the upside-down double well potential.
Our choice of basis functions might also suppress the `decay' of the
states.
(3) The normal-ordering problem becomes serious for weaker couplings.
Our previous calculation\cite{HOTsix} shows that the meson mass
depends
on the fermion mass almost linearly for a wide range of the fermion
mass.
But the equation (\ref{odd1}) shows, according to the variational
principle,
that the meson mass must be smaller than $\tilde\mu=\sqrt{1+4\pi
cm}$, which
has the linear dependence on the fermion mass only in the strong
coupling
regions. For weaker fermion masses, this appears to put a stringent
``upper bound.'' Of course this should not be true. (The
structureless
one-boson approximation to the meson cannot be better than that
including
up to fermion six-body states, especially for weak couplings.)
The ``coupling'' $c$ must be renormalized and must have a nontrivial
fermion mass dependence. The investigation in this direction is now
in progress\cite{HOTrenorm}. For the purpose of the present paper,
however,
it is sufficient to notice that the renormalization effects are
negligibly
small for small fermion masses. The fermion mass dependence is almost
identical to that of our previous calculation.

\begin{figure}
\epsfysize=0.42\vsize
\epsfbox{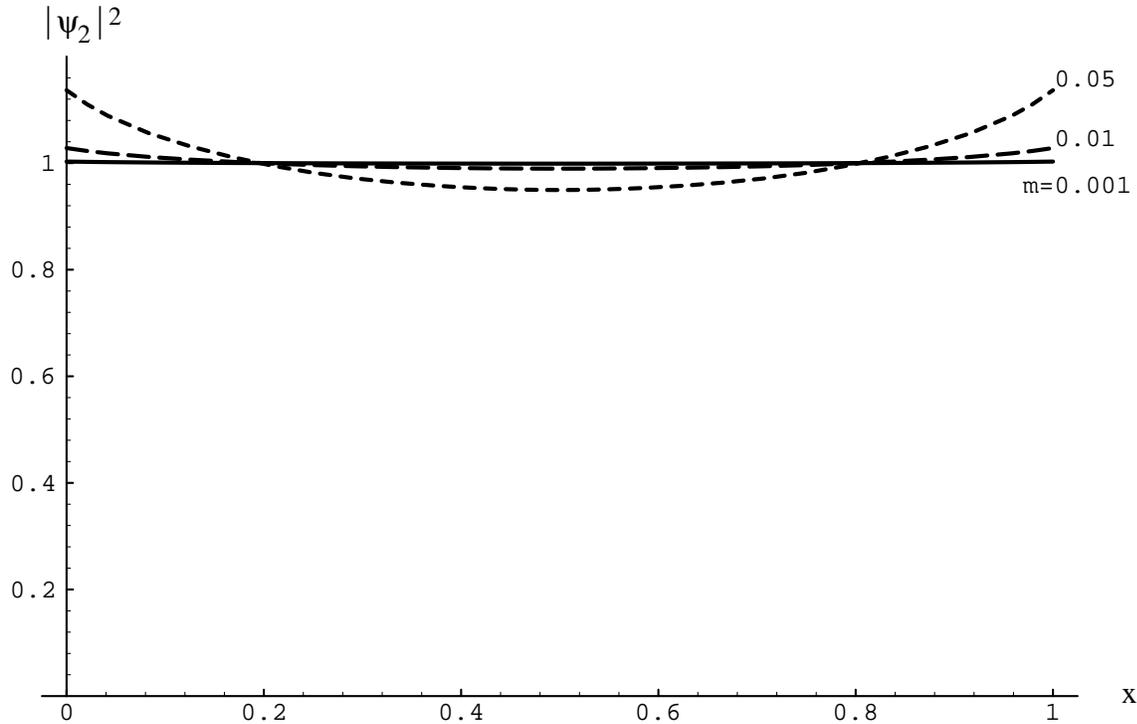}
\caption{Fermion two-body wave function of the meson,
$\vert\psi_2(x,1-x)\vert^2$ is shown for various values of the
fermion mass,
where $\psi_2\equiv\phi_1+\psi_2^{(3)}$.
This should be compared with Fig.~5 of our previous work.}
\label{mesonwavefunc}
\end{figure}

Another pathology of the bosonized model can be seen
in the wave function of the
meson state.
It is known that the fermion two-body wave function of the
meson state, $\psi_2(x,1-x)\equiv\phi_1+\psi_2^{(3)}(x,1-x)$,
must vanish at $x=0$ and $x=1$ in the massive theory\cite{bergknoff}.
The calculated wave function shown in Fig.~\ref{mesonwavefunc}
does not satisfy
this requirement. Furthermore, it is not a concave function of $x$.
(See Fig.~5 of Ref.~\cite{HOTsix}, for example.)
We do not know why it does not
have the correct behavior.

It would be interesting to use the bosonized theory to investigate
the
effects of the theta vacuum~\cite{C}, though the investigation in the
original fermionic theory requires an intensive
study~\cite{HOTtheta}.


\acknowledgments

The authors are grateful to colleagues for many discussions. One of
the
authors (K.H.) would like to thank Robert J. Perry for interesting
discussions, especially on the renormalization of the coupling due to
normal-ordering,
and the hospitality in the Ohio-State University. This work is
supported by a Grant-in-Aid for Scientific Research form the Ministry
of
Education, Science and Culture of Japan (No.06640404).
%
%

%
%

\end{document}